\documentclass[twocolumn]{IECEjournal} % lineno, proof, onecolumn

\usepackage{hyperref, xcolor, url, multirow, adjustbox, graphicx, threeparttable, diagbox, tabularx, changepage, amsmath, amssymb, cleveref, hyphenat, enumitem, lipsum, multicol, float, newfloat, array, pifont, balance, tikz, forest}

\usepackage[ruled]{algorithm2e}

\usetikzlibrary{arrows.meta}
\forestset{edges}
\graphicspath{{res/},{media/},{figures/}}

\definecolor{darkpurple}{RGB}{104, 53, 155}  
\definecolor{lawngreen}{rgb}{0.51, 0.97, 0.02}  
\definecolor{pink}{rgb}{1, 0.02, 0.48}  
\definecolor{airforce}{rgb}{0.35, 0.56, 0.64}  
\definecolor{hidden-draw}{RGB}{2, 2, 2}  
\definecolor{hidden-pink}{rgb}{0.97, 0.93, 0.77}  
\definecolor{level0}{rgb}{0.66, 0.86, 0.71}  
\definecolor{level1}{rgb}{0.97, 0.90, 0.83}  
\definecolor{level2}{rgb}{0.81, 0.79, 0.99}  
\definecolor{level3}{rgb}{1.01, 0.72, 0.74}  
\definecolor{level4}{rgb}{0.48, 0.98, 0.02}

\articletype{Review}
\submitdate{01 August 2024}
\acceptdate{accept-date}
\pubdate{pub-date}

\jnlsetup{%
    author  = {Wang, P., Zheng, H., Dai, S., Wang, Y., Gu, X., Wu, Y, &  Wang, X.},
    journal = {Chinese Journal of Information Fusion},
    title   = {A Survey of Spatio-Temporal EEG data Analysis: from Models to Applications},  % Manuscript title
    ctitle  = {A Survey of Spatio-Temporal EEG data Analysis: from Models to Applications},  % Manuscript title
    volume  = {1},
    number  = {1},
    year    = {2024},
    doi     = {10.62762/IECE.2024.000000}
}
\dataavail{%
The data used to support the findings of this study are available from the corresponding author upon request.}

\makeatother
\begin{document}
%% -- Complete author and affiliation info
% \author[<superscipt>][<corresponding.author.email>]{<name>}[<orcid>]
\author[1]{Pengfei Wang}[0009-0008-5593-2653]
\author[1]{Huanran Zheng}[0009-0005-6261-6702]
\author[1]{Silong Dai}[0009-0002-7065-6059]
\author[1]{Yiqiao Wang}[0009-0007-2344-6055]
\author[1]{Xiaotian Gu}[0009-0002-8406-7511]
\author[1][ybwu@cs.ecnu.edu.cn]{Yuanbin Wu}[0009-0009-5670-4940]
\author[1]{Xiaoling Wang}[0000-0002-4594-6946]

\affil[1]{School of Computer Science and Technology, East China Normal University, Shanghai 200062, China}
% \affil[2]{School of B, University of B, Spokane 02139, USA}

%\thanks[$\dagger$]{These authors contributed equally to this work}

\ifproof
\input{proof}
\fi

\maketitle

\abstract{In recent years, the field of electroencephalography (EEG) analysis has witnessed remarkable advancements, driven by the integration of machine learning and artificial intelligence. This survey aims to encapsulate the latest developments, focusing on emerging methods and technologies that are poised to transform our comprehension and interpretation of brain activity. We delve into self-supervised learning methods that enable the robust representation of brain signals, which are fundamental for a variety of downstream applications. We also explore emerging discriminative methods, including graph neural networks (GNN), foundation models, and large language models (LLMs)-based approaches. Furthermore, we examine generative technologies that harness EEG data to produce images or text, offering novel perspectives on brain activity visualization and interpretation. The survey provides an extensive overview of these cutting-edge techniques, their current applications, and the profound implications they hold for future research and clinical practice. The relevant literature and open-source materials have been compiled and are consistently being refreshed at \url{https://github.com/wpf535236337/LLMs4TS}}


%% -- citationblock
% \citationblock
\section{Introduction}\label{sec:Introduction}
% Background: Briefly introduce EEG analysis, its importance, and the recent integration of machine learning and artificial intelligence.
Electroencephalography (EEG) has long been a cornerstone in the study of brain function, offering a non-invasive means to monitor electrical activity within the brain. Non-invasive are easier to implement without surgery, but they lack simultaneous consideration of temporal and spatial resolution, as well as the ability to capture deep brain information. In contrast, invasive methods like Stereoelectroencephalography (SEEG)\cite{david2011imaging} can measure these brain signals more precise with higher signal-to-noise data \cite{cai2023mbrain}, albeit requiring surgical procedures to insert recording devices. Overall, non-invasive signals are relatively safer, more portable, have greater potential for use, and are applicable to a wider population, reflecting voltage fluctuations caused by ion currents in neurons. 

%challenge
% EEG data, as a crucial temporal signal, provides essential technical support for enhancing our exploration of brain functions and improving disease diagnosis. The application of deep learning techniques\cite{craik2019deep} to EEG data analysis plays a significant role in advancing brain-computer interface technology. As our understanding of the brain deepens and computational methods advance, the field of EEG analysis has reached a pivotal moment. However, the high dimensionality, noise, and nonlinear dynamics of EEG data present significant challenges in extracting meaningful information. Traditional analysis\cite{hosseini2020review} methods struggle to cope with the complexities of brain activity, highlighting the need for more sophisticated and advanced techniques to fully harness the potential of this technology.
While our understanding of the brain deepens and computational methods advance\cite{craik2019deep, hosseini2020review}, the field of EEG analysis faces many challenges. The first challenge is the effective capture of representations in EEG data, particularly in the absence of labels. The second challenge involves the identification and classification of complex and subtle patterns within brain activity, requiring advanced discriminative methods that can accurately interpret the nuanced differences indicative of various brain states or conditions. Lastly, the challenge of creating meaningful visualizations or interpretations from EEG data calls for generative methods that can transform the abstract EEG signals into more tangible and comprehensible forms, such as images or text, thereby enhancing our understanding of the brain's intricate workings. Addressing these challenges collectively advances the field of EEG analysis, making it more robust, insightful, and applicable to a wider range of scientific and clinical applications.

% Motivation: Explain the need for a comprehensive survey on the latest advancements.
% advancements
In response to aforementioned challenges, recent developments in deep learning and artificial intelligence have paved the way for more robust and nuanced EEG analysis strategies. This paper surveys three key areas of advancement that are reshaping the field of EEG analysis:
\begin{itemize}
    \item \textbf{Representation Learning in EEG Analysis}: Representation learning is the first fundamental step in EEG analysis, concentrate on automatically extracting useful features from EEG signals. Self-supervised learning methods are being employed to develop robust signal representations that enhance the precision and interpretability of downstream tasks. These unsupervised learning methods are naturally suited for the vast amounts of brain signal data and mimic human learning processes.
    \item \textbf{Discriminative EEG Analysis}: Discriminative methods focus on distinguishing between different categories or patterns in EEG signals. Advanced architectures such as Graph Neural Networks (GNNs), Foundation Models, and LLMs-based Methods are being utilized to gain deeper insights into brain activity. These architectures efficiently capture discriminative patterns, which are crucial for understanding complex neural processes.
    \item \textbf{Generative EEG Analysis}:
    Generative methods aim to generate new modalities or signal data from EEG signals. Innovative approaches such as diffusion  produce images or text from EEG data are providing novel approaches to the understanding and visualization of brain activity. These generative techniques are also important applications for AI-generated content (AIGC).
\end{itemize}

% Objectives
This paper aims to provide a comprehensive overview of these cutting-edge techniques, discuss their details, and consider the significant implications they hold for future research and clinical practice in EEG analysis.
% Structure of the Paper
% The remainder of this paper is organized as follows: Section \ref{sec:ssl} discusses the robust representation learning strategy and its significance in EEG data analysis. Section \ref{sec:survey} summarize the background and related surveys of our work. Section \ref{sec:architecture} explores the emerging discriminative  architecture, detailing the role of GNNs \ref{subsec:gnn}, Foundation Models \ref{subsec:foundation_model}, and LLMs-based Methods \ref{subsec:llms}. Section \ref{sec:generative} examines the innovative generative applications of EEG data. Section 
%  \ref{sec:dataset_and_metric} provides an overview of the most widely used datasets and the key metrics employed to assess the performance of various EEG analysis models. Finally, Section \ref{sec:conclusion} concludes the paper and discusses the future directions for EEG analysis.
The remainder of this paper is organized as follows:  Section \ref{sec:survey} summarizes the background and related surveys of our work. Section \ref{sec:ssl} discusses the robust representation learning strategy and its significance in EEG data analysis. Section \ref{sec:architecture} explores the emergent discriminative architecture, detailing the role of GNNs (\ref{subsec:gnn}), Foundation Models (\ref{subsec:foundation_model}), and LLMs-based Methods (\ref{subsec:llms}). Section \ref{sec:generative} examines the innovative generative applications of EEG data. Section \ref{sec:dataset_and_metric} provides an introduction of the most widely used datasets and the key metrics employed to assess the performance of various EEG analysis models. Finally, Section \ref{sec:conclusion} concludes the paper and discusses potential future directions for EEG analysis.
\tikzstyle{my-box}=[
    rectangle,
    draw=hidden-draw,
    rounded corners,
    text opacity=1,
    minimum height=1.5em,
    minimum width=5em,
    inner sep=2pt,
    align=center,
    fill opacity=.5,
    line width=0.8pt,
]
\tikzstyle{leaf}=[my-box, minimum height=1.5em,
    fill=hidden-pink!80, text=black, align=left, font=\normalsize,
    inner xsep=2pt,
    inner ysep=4pt,
    line width=0.8pt,
]
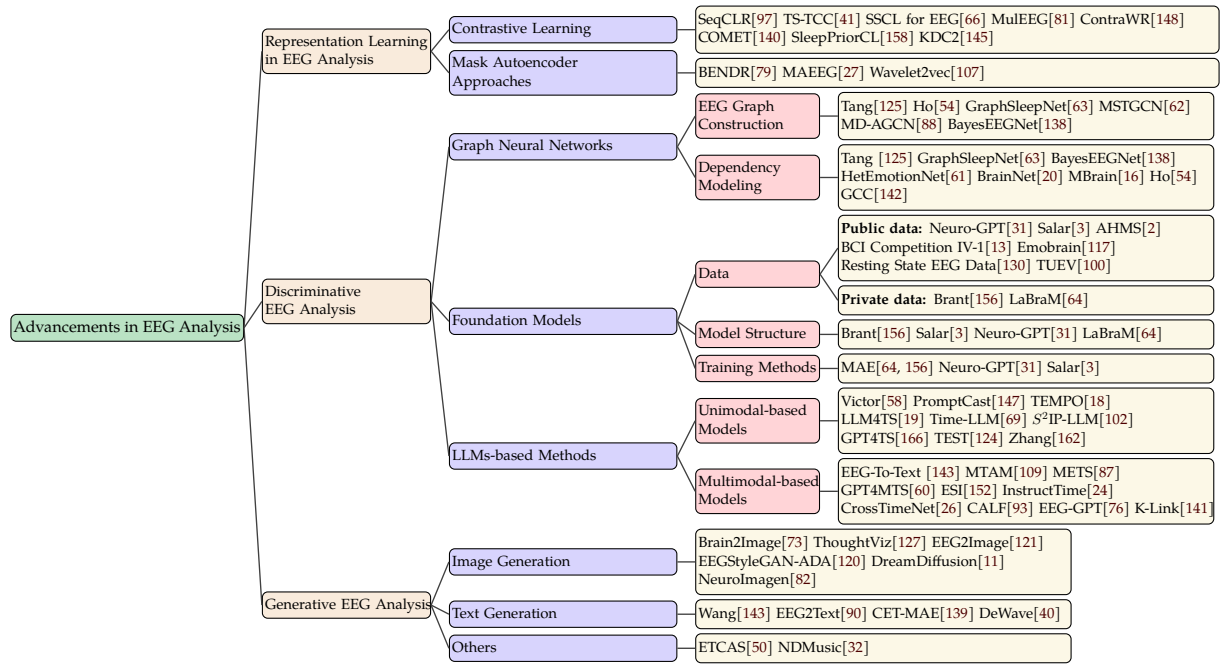
\begin{figure*}[!t]
    \centering
    \resizebox{0.87\textwidth}{!}
    {  
        \begin{forest}
            for tree={
                fill=level0!80,
                grow=east,
                reversed=true,
                anchor=base west,
                parent anchor=east,
                child anchor=west,
                base=left,
                font=\large,
                rectangle,
                draw=hidden-draw,
                rounded corners,
                align=left,
                minimum width=4em,
                edge+={darkgray, line width=1pt},
                s sep=3pt,
                inner xsep=2pt,
                inner ysep=3pt,
                line width=0.8pt,
                ver/.style={rotate=90, child anchor=north, parent anchor=south, anchor=center},
            },
            where level=1{text width=11em,font=\normalsize,fill=level1!80,}{} ,
            where level=2{text width=15em,font=\normalsize,fill=level2!80,}{} ,
            where level=3{text width=8em,font=\normalsize,fill=level3!60,}{} ,
            where level=4{text width=10em,font=\normalsize,fill=level4!20,}{} ,
            where level=5{text width=5em,font=\normalsize,fill=level4!20,}{} ,
            [Advancements in EEG Analysis
                [Representation Learning \\ in EEG Analysis
                    [Contrastive Learning
                        [SeqCLR\cite{mohsenvand2020contrastive} TS-TCC\cite{eldele2021time} SSCL for EEG\cite{jiang2021self} MulEEG\cite{kumar2022muleeg} ContraWR\cite{yang2023self} \\COMET\cite{wang2024contrast} SleepPriorCL\cite{zhang2021sleeppriorcl} KDC2\cite{weng2023knowledge}, leaf, text width=35em]
                    ]
                    [Mask Autoencoder \\ Approaches
                        [BENDR\cite{kostas2021bendr} MAEEG\cite{chien2022maeeg} Wavelet2vec\cite{peng2023wavelet2vec}, leaf, text width=35em]   
                    ]    
                ]
                [Discriminative \\ EEG Analysis
                    [Graph Neural Networks
                        [EEG Graph \\ Construction
                            [Tang\cite{Tang2021SelfSupervisedGN} Ho\cite{Ho2022SelfSupervisedLF} GraphSleepNet\cite{jia2020graphsleepnet} MSTGCN\cite{Jia2021MSTGCN}\\ MD-AGCN\cite{Li2021AMA} BayesEEGNet\cite{Wang2023ABG}, leaf, text width=25em]    
                        ]
                        [Dependency \\ Modeling
                            [Tang \cite{Tang2021SelfSupervisedGN} GraphSleepNet\cite{jia2020graphsleepnet} BayesEEGNet\cite{Wang2023ABG}\\ HetEmotionNet\cite{Jia2021HetEmotionNet} BrainNet\cite{Chen2022BrainNet} MBrain\cite{cai2023mbrain} Ho\cite{Ho2022SelfSupervisedLF}\\ GCC\cite{Wang2023GraphAwareCF}, leaf, text width=25em] 
                        ]
                    ]
                    [Foundation Models
                        [Data
                            [\textbf{Public data:} Neuro-GPT\cite{Cui2023NeuroGPTTA} Salar\cite{Abbaspourazad2023LargescaleTO} AHMS\cite{2019AppleH}\\ BCI Competition IV-1\cite{Blankertz2007TheNB} Emobrain\cite{Savran2006EmotionDI} \\Resting State EEG Data\cite{Trujillo2017TheEO} TUEV\cite{Obeid2016TheTU}, leaf, text width=25em]
                            [\textbf{Private data:} Brant\cite{Zhang2023BrantFM} LaBraM\cite{Jiang2024LargeBM}, leaf, text width=25em]
                        ]
                        [Model Structure
                            [Brant\cite{Zhang2023BrantFM} Salar\cite{Abbaspourazad2023LargescaleTO} Neuro-GPT\cite{Cui2023NeuroGPTTA} LaBraM\cite{Jiang2024LargeBM}, leaf, text width=25em]
                        ]
                        [Training Methods
                            [MAE\cite{Zhang2023BrantFM,Jiang2024LargeBM} Neuro-GPT\cite{Cui2023NeuroGPTTA} Salar\cite{Abbaspourazad2023LargescaleTO}, leaf, text width=25em]
                        ]
                    ]
                    [LLMs-based Methods
                        [Unimodal-based \\ Models
                            [Victor\cite{iapascurta2023nlp} PromptCast\cite{xue2023promptcast} TEMPO\cite{cao2023tempo} \\LLM4TS\cite{chang2023llm4ts} Time-LLM\cite{jin2023time} $S^2$IP-LLM\cite{pan2024textbf} \\GPT4TS\cite{zhou2023one} TEST\cite{sun2023test} Zhang\cite{zhang2024word}, leaf, text width=25em]
                        ]
                        [Multimodal-based \\ Models
                            [EEG-To-Text \cite{wang2022open} MTAM\cite{qiu2023can} METS\cite{li2024frozen} \\GPT4MTS\cite{jia2024gpt4mts} ESI\cite{yu2024ecg} InstructTime\cite{cheng2024advancing} \\CrossTimeNet\cite{cheng2024learning} CALF\cite{liu2024taming} EEG-GPT\cite{kim2024eeg} K-Link\cite{wang2024k}, leaf, text width=25em]
                        ]
                    ]
                ]
                [Generative EEG Analysis
                    [Image Generation
                        [Brain2Image\cite{kavasidis2017brain2image} ThoughtViz\cite{tirupattur2018thoughtviz} EEG2Image\cite{singh2023eeg2image} \\EEGStyleGAN-ADA\cite{singh2024learning} DreamDiffusion\cite{bai2023dreamdiffusion} \\NeuroImagen\cite{lan2023seeing}, leaf, text width=25em]
                    ]
                    [Text Generation
                        [Wang\cite{wang2022open} EEG2Text\cite{liu2024eeg2text} CET-MAE\cite{wang2024enhancing} DeWave\cite{duan2024dewave}, leaf, text width=25em]
                    ]
                    [Others
                        [ETCAS\cite{guo2023end} NDMusic\cite{daly2023neural}, leaf, text width=25em]
                    ]
                ]
            ]
        \end{forest}
    }
% \vspace{-0.5em}
\caption{A comprehensive taxonomy of advancements in EEG Analysis}
% \vspace{-1em}
\label{fig:taxonomy}
\end{figure*}

\section{Related survey}\label{sec:survey}
% 
% Summarize major reviews and surveys, highlighting their focus areas.
%  Identify the areas not fully covered by previous surveys that this paper addresses
\subsection{Existing Surveys on EEG Analysis}
In the domain of EEG-related concepts and research, numerous review studies have provided comprehensive summaries. Hosseini et al. \cite{hosseini2020review} introduced the application of machine learning in EEG signal processing, covering traditional methods such as Support Vector Machines (SVM), k-Nearest Neighbors (kNN), and Naive Bayes in classification scenarios. However, this review did not consider the extensive discussion of deep learning algorithms that have demonstrated superior performance. Jiang et al. \cite{Jiang2019artifacts} discussed the removal of artifacts from EEG signals, making their review more detailed in technical aspects. Nevertheless, their work did not cover deep learning algorithms and did not consider a broader range of EEG downstream tasks. In contrast, Zhang et al. \cite{Zhang2019ASO} provided a more comprehensive perspective, introducing the origins and applications of Brain-Computer Interface (BCI) and discussing the integration of mainstream deep learning algorithms such as Convolutional Neural Networks (CNN), Recurrent Neural Networks (RNN), and Generative Adversarial Networks (GAN) with EEG tasks. With the continuous innovation in artificial intelligence community, EEG research based on foundational models and large language models has begun to emerge. However, to the best of our knowledge, there is currently no literature that reviews EEG analysis from a more holistic frontier technology perspective, which is the gap this paper aims to fill.

\subsection{Emerging Surveys on General Time-Series Analysis}
In the general time series domain, a substantial body of work has summarized the application of the latest technologies in various downstream tasks. Zhang et al.\cite{Zhang2023SelfSupervisedLF} categorized existing self-supervised learning-based time series analysis methods into three types: generative, contrastive, and adversarial, and discussed their key intuitions and main frameworks in detail. Jin et al.\cite{Jin2023GNN4TSsurvey} provided an overview of the application of graph neural networks in time series tasks such as forecasting, classification, imputation, and anomaly detection. Liang et al.\cite{liang2024foundation} reviewed foundational models in time series analysis from the perspectives of model architectures, pre-training techniques, adaptation methods, and data modalities. Similarly, \cite{Jiang2024LargeBM, zhang2024large, jin2023large} systematically outlined methods and procedures for time series analysis based on large language models. Yang et al. \cite{yang2024survey} reviewed the application of diffusion models in time series and spatio-temporal data. Additionally, there are some works focusing on more specific model architectures or downstream tasks\cite{Zhang2024LargeLM, ijcai2023survey}. We refer the reader to the corresponding publication for a more in-depth understanding.

% Although there are many reviews in the general time series domain, few surveys focus solely on EEG data. Meanwhile, EEG data has some unique characteristics, and a significant amount of related work has emerged in recent years, warranting a thorough review and summary. Therefore, this paper aims to provide a comprehensive overview of the cutting-edge techniques, discuss their details, and consider the significant implications they hold for future research and clinical practice in EEG analysis.
Although numerous reviews exist within the broader time series field, few surveys concentrate exclusively on EEG data. Moreover, EEG data possesses unique characteristics, and a substantial body of related work has emerged recently. Thus necessitating a comprehensive review and synthesis, this paper seeks to offer an in-depth examination of state-of-the-art techniques, elaborate on their intricacies, and explore their profound implications for future EEG research and clinical applications.

\section{Representation Learning in EEG Analysis}\label{sec:ssl}
% \section{Self-Supervised Learning}
In recent years, deep learning has excelled in extracting hidden patterns and features of the data. Typically, feature extraction models based on deep learning rely heavily on large volumes of labeled data, a method commonly referred to as supervised learning. However, in certain practical applications, particularly in time-series data such as Electroencephalograms (EEG), acquiring extensive labeled data is both time-consuming and costly. As an alternative, Self-Supervised Learning (SSL) has garnered increasing attention due to its label efficiency and generalization capabilities. SSL, a subset of unsupervised learning, extracts supervisory signals by solving tasks automatically generated from unlabeled data, thereby creating valuable representations for downstream tasks.

With the significant success of SSL in fields such as computer Vision(CV)\cite{liu2023fac} and Natural Language Processing(NLP)\cite{gao2021simcse}, its application to time-series data appears particularly promising. However, directly applying tasks designed for visual or linguistic processing to time-series data is challenging and often yields limited effectiveness. The primary reasons include:

\begin{itemize}

\item Time-series data possess unique attributes such as seasonality, trends, and frequency domain information, which are typically not considered in tasks designed for images or language.

\item Common data augmentation techniques in computer vision, such as rotation, flipping, and cropping, can disrupt the temporal dependencies and integrity of time-series data, such as EEG signals. For instance, rotating or flipping the time points in an EEG signal could completely lose physiological significance and contextual information.

\item Many time-series datasets are multidimensional, with each dimension potentially representing a different measurement channel. This contrasts with handling single images or text data, requiring synchronous analysis and processing across multiple dimensions.

\end{itemize}

To address these issues, this section summarizes two main paradigms of SSL: contrastive learning, which trains models to distinguish between similar and dissimilar pairs of data points and masked autoencoders, which aim to learn the intrinsic feature information of the data. All of the methods are summarized in \cref{tab:SSL}.

\subsection{Contrastive Learning}
Contrastive learning is a self-supervised learning method that acquires invariant representations of data by learning the similarities and differences between samples. This approach maps similar samples to proximate representation spaces and dissimilar samples to distant ones, thereby enabling the learning of generalized feature representations without the need for explicit label information. Formally, given a set of samples $\mathcal{X}=\left\{x^{1},x^{2},\cdots,x^{N}\right\}$, contrastive learning aims to learn a mapping function $f$ that maximizes the similarity between positive sample pairs of the same class and minimizes the similarity between negative sample pairs of different classes. For positive sample pairs $(x,x^{+})$ and negative sample pairs $(x,x^{-})$, the objective of contrastive learning is to optimize the following loss function:
\begin{equation}
    L\left(x,x^+,x^-\right)=-\log\left(\frac{e^{f\left(x,x^+\right)/\tau}}{e^{f(x,x^+)/\tau}+e^{f(x,x^-)/\tau}}\right)
\end{equation}
where $f(x,x^+)$ denotes the similarity of feature representations for positive pairs, $f(x,x^-)$ for negative pairs, and $\tau$ is a temperature parameter that adjusts the scale of similarity. The intuitive interpretation of this loss function is that by maximizing the similarity of positive pairs while minimizing that of negative pairs, the model learns high-level semantic relationships between samples, resulting in more distinctive representations.

Contrastive learning offers significant advantages in the analysis of multivariate medical signals. Firstly, as an unsupervised learning method, it does not require explicit labeling, thus providing an efficient training approach for the domain of multivariate medical signals where labeling is difficult and costly. Secondly, by learning the similarities and differences between positive and negative sample pairs, contrastive learning can capture more generalizable feature representations. This is particularly applicable to multivariate medical signals collected from different subjects or devices, enabling the learning of core discriminative representations and reducing the impact of domain shift.

\subsubsection{Based on Data Augmentation}

Data augmentation is an indispensable component of contrastive learning. It generates different views of input samples using data augmentation techniques, and then learns representations by maximizing the similarity between views of the same sample while minimizing the similarity between views of different samples. SeqCLR\cite{mohsenvand2020contrastive} introduces a set of data augmentation techniques specifically for EEG and extends the SimCLR \cite{chen2020simple} framework to extract channel-level features from EEG data.

TS-TCC \cite{eldele2021time} generates different views of input data using both strong and weak augmentation methods. Weak augmentation employs jittering and scaling strategies, while strong augmentation uses permutation and jittering strategies, applying them to the temporal contrast module of EEG signals for temporal representation learning. This method maximizes the similarity between contexts of the same sample while minimizing the similarity between contexts of different samples. Jiang et al.\cite{jiang2021self} applies transformations such as horizontal flipping and adding Gaussian noise to EEG signals, then learns the correlation between signals by measuring the feature similarity of these transformed signal pairs. Additionally, the authors explore the impact of transformation combinations on the network's representation capability to find the optimal combination for downstream tasks. mulEEG \cite{kumar2022muleeg} proposes a novel multi-view self-supervised method. By designing EEG augmentation strategies and introducing a diversity loss function, mulEEG effectively leverages complementary information from multiple views to learn better representations. However, these EEG data augmentation methods often lead to sampling bias\cite{chuang2020debiased}, especially for noisy EEG data, which can significantly affect performance\cite{robinson2020contrastive}. To address these limitations, ContraWR \cite{yang2023self} constructs positive sample pairs using data augmentation and employs global average representations as negative samples to provide contrastive information, thereby learning robust EEG representations without labels. Additionally, ContraWR assigns greater weight to closer samples when calculating the global average.

Existing contrastive learning methods primarily focus on a single data level and fail to fully exploit the complexity of EEG signals. Therefore, COMET \cite{wang2024contrast} leverages all data levels of medical time-series, including patient, trial, sample, and observation levels, to design a hierarchical contrastive representation learning framework. Its advantage lies in fully utilizing the hierarchical structure of medical time-series, enabling a more comprehensive understanding of the intrinsic relationships within the data.

\subsubsection{Combined with Expert Knowledge}

Expert knowledge contrastive learning is a relatively new representation learning framework. Generally, this modeling framework incorporates expert prior knowledge or information into deep neural networks to guide model training. In a contrastive learning framework, prior knowledge can help the model select the correct positive and negative samples during training.SleepPriorCL \cite{zhang2021sleeppriorcl} was proposed to mitigate the sampling bias problem in data augmentation-based contrastive learning. It is well known that each sleep stage occupies a certain frequency range. The authors utilized this fact to calculate the energy of these frequency bands and used it as prior knowledge for training. Specifically, the authors calculated the rhythm energy vector $E=[E(\delta),E(\theta),E(\alpha),E(\beta)]$ for each EEG segment $x$, referred to as prior features, and then defined the dissimilarity $d_{i,j}$ between the anchor $x_i$ and the sample $x_j$ as follows:

\begin{equation}
    d_{i,j}=\log\left(\left\|E_i-E_j\right\|_2\right)
\end{equation}

Samples are ranked by dissimilarity, with the top K samples selected as positive samples and the rest as negative samples. Additionally, SleepPriorCL introduces a mechanism to adjust the gradient penalty strength of each sample based on its confidence as a positive or negative sample. To achieve this, each sample is assigned a customized temperature. The multi-positive contrastive loss is modified as follows:
% \scalebox{0.8}{
% \begin{equation}
%     \mathcal{L}\left(x_i\right)=\frac{-1}{\left|P(i)\right|}\sum_{p\in P(i)}\log\frac{\exp\left(s_{i,p}/\tau_p\right)}{\exp\left(s_{i,p}/\tau_p\right)+\sum_{n\in N(i)}\exp\left(s_{i,n}/\tau_n\right)}
% \end{equation}
% }

% https://zhuanlan.zhihu.com/p/598797769
\begin{equation}
    \scriptstyle{\mathcal{L}(x_i) = \frac{-1}{|P(i)|} \sum_{p \in P(i)} \log \left( \frac{\exp\left(\frac{s_{i,p}}{\tau_p}\right)}{\exp\left(\frac{s_{i,p}}{\tau_p}\right) + \sum_{n \in N(i)} \exp\left(\frac{s_{i,n}}{\tau_n}\right)} \right)}
\end{equation}

where $x_i$ is the sleep epoch, $s_{i,j}$ is the cosine similarity between $z_i$ and $z_j$, and $z_i$ and $z_j$ are the vectors of $x_i$ after encoding and projection. The index $i$ is referred to as the anchor, the index $p$ as the positive sample, $N(i)$ is the set of all negative samples in the batch, and the index $n$ as the negative sample. $P(i)$ is the set of positive samples containing all true positive samples of $x_i$ in the batch.

KDC2 \cite{weng2023knowledge} is based on the neural theory of EEG generation, which states that EEG signals are produced by synchronized synaptic activity that stimulates neuronal excitation, generating a negative extracellular voltage that transforms neurons into dipoles. The voltage generated by the dipoles is transmitted to the scalp via capacitive and volume conduction and is captured by electrodes as EEG signals. Therefore, the authors constructed scalp and neural views to describe the external and internal information of brain activity, respectively, and designed a knowledge-driven cross-view contrastive loss to extract neural knowledge by contrasting the same augmented samples between views. Positive sample pairs are composed of representations of the same augmented samples in different views, while negative sample pairs are composed of representations of different augmented samples in different views. By minimizing the distance between positive sample pairs and maximizing the distance between negative sample pairs, the model learns complementary features that describe the internal and external manifestations of brain activity. The designed cross-view contrastive loss can be calculated as follows:

\begin{equation}
\mathcal{L}_{cross}=-\frac{1}{|\mathcal{B}|}log(\frac{pair^{+}}{pair^{+}+pair^{-}})
\end{equation}

\begin{equation}
pair^{+}=\sum_{b\in\mathcal{B}}\sum_{i=0}^{m}exp(s(r_{sa,b}^{i},r_{ta,b}^{i})/\tau)
\end{equation}

\begin{equation}
pair^{-}=\sum_{b\in\mathcal{B}}\sum_{i=0}^{m}\sum_{j=i+1}^{m}exp(s(r_{sa,b}^{i},r_{ta,b}^{j})/\tau) 
\end{equation}
where $pair^{+}$ and $pair^{-}$ represent the cross-view positive and negative pairs, respectively, $\mathcal{B}$ is the sample batch, and $\tau$ is the temperature parameter. The function $s()$ represents the cosine similarity. The representation generated from the scalp view is denoted as $r_{s}$, and the representation generated from the inner neural topology view is denoted as $r_{t}$. $r_{sa}$ and $r_{ta}$ represent the corresponding augmented samples, and $b$ indexes the samples contained in the batch.

\subsection{Mask Autoencoder Approaches}

Masked language modeling is a widely adopted method for pre-training in NLP. BERT \cite{devlin2018bert} retains a portion of the input sequence and predicts the missing content during the training phase, which generates effective representations for various downstream tasks. MAE can be represented as: 
\begin{equation}
    x_{m}=\mathcal{M}(x),\quad z=E(x_{m}),\quad\tilde{x}=D(z),
\end{equation}
\begin{equation}
    \mathcal{L}=\mathcal{M}(\|x-\tilde{x}\|_{2})
\end{equation}
where $\mathcal{M}(\cdot)$ denotes the masking operation, $x_{m}$ represents the masked input, $E(\cdot)$ and $D(\cdot)$ represent the encoder and decoder.

Inspired by this, BENDR \cite{kostas2021bendr} follows the wav2vec2.0 \cite{baevski2020wav2vec} architecture. It first encodes EEG data into temporal embeddings using 1D convolutions, then creates a mask vector to randomly mask these embeddings. A transformer-based module \cite{vaswani2017attention} is then used to extract temporal correlations and output the reconstructed embeddings. The contrastive loss function aims to make the reconstructed embeddings as similar as possible to the original unmasked embeddings while making them as different as possible from the remaining embeddings. It can be calculated as follows:

\begin{equation}
    \mathcal{L}=-log\frac{exp(cossim(c_t,b_t))/\kappa}{\sum_{b_i\in B_D}\exp(cossim(c_t,b_i))/\kappa}
\end{equation}
where $c_t$ represents the output of the transformer module at position $t$, $b_i$ represents the original vector at some offset $i$, $B_D$ is a set of 20 negative samples uniformly selected from the same sequence, along with $b_t$, $cossim$ denotes the cosine similarity, and $\kappa$ is a temperature parameter controlling the contrastive loss.

MAEEG \cite{chien2022maeeg} has a similar structure to BENDR but includes two additional layers to map the output of the transformer module back to the original EEG dimensions. The reconstruction loss is calculated by comparing the reconstructed EEG $(\hat{\mathrm{x}})$ with the input EEG $(\mathrm{x})$ signal, using the formula $1-\frac{\mathbf{\hat{x}}\cdot\mathbf{x}}{\|\mathbf{\hat{x}}\|\|\mathbf{x}\|}$. The key difference between BENDR and MAEEG is that MAEEG learns representations by minimizing the reconstruction loss rather than using contrastive learning.

Unlike the above two methods that mask temporal embeddings, WAVELET2VEC \cite{peng2023wavelet2vec} performs masking and reconstruction tasks in different frequency bands to capture time-frequency information. Specifically, the authors apply low-pass and high-pass filtering to the raw EEG signal, recursively calculate the coefficients of each level of decomposition, and obtain wavelets in different frequency bands. They then design an encoder consisting of six parallel ViT   \cite{dosovitskiy2020image} units, each corresponding to a frequency band wavelet. Each wavelet is flattened and divided into patches, and 10\% of the input patches are randomly masked. The decoder reconstructs the missing patch sequences, and self-supervised pre-training is performed by minimizing the Euclidean distance between the patch sequences of the original signal and the reconstructed patch sequences. This method forces the model to learn the time-frequency information and understand its correlations by masking the frequency patch sequences of the EEG.

\begin{table*}[htb]
    \centering
    \caption{Summary of Self-Supervised Learning for EEG Analysis}
    \resizebox{\textwidth}{!}{
    \begin{tabular}{ccccccc}
    \toprule[1pt]
       \textbf{Modality} & \textbf{Method} & \textbf{Strategy}  & \textbf{Backbone} & \textbf{Task}  & \textbf{Datasets} & \textbf{Metric} \\
       \hline
        \multirow{8}{*}{Contrastive Learning}  
        & SeqCLR\cite{mohsenvand2020contrastive}  & Signal transformation & CNN \& GRU & Multiple tasks  & THU\cite{Obeid2016TheTU}, SEED\cite{zheng2017identifying}, SleepEDF\cite{kemp2000analysis}, ISRUC-S3\cite{KHALIGHI2016180}  & Accuracy\\
        & TS-TCC\cite{eldele2021time}   & Weak \& strong augmentation  & Transformer & Sleep \& seizure detection &HAR\cite{Anguita2013APD}, SleepEDF\cite{kemp2000analysis}, ESR\cite{Andrzejak2001IndicationsON},
        FD\cite{lessmeier2016condition} & Accuracy,
        F1 \\
        & SSCL for EEG\cite{jiang2021self}    & Signal transformation & CNN & Sleep stage classification &SleepEDF\cite{kemp2000analysis}, DOD\cite{guillot2020dreem} & Accuracy, F1\\
        & MulEEG\cite{kumar2022muleeg}  & Multi-view contrast & CNN & Sleep stage classification &SleepEDF\cite{kemp2000analysis}, SHHS\cite{zhang2018national} &Accuracy, Kappa, F1 \\
        & ContraWR\cite{yang2023self}  & Non-negative contrast& CNN & Sleep stage classification &SHHS\cite{zhang2018national}, SleepEDF\cite{kemp2000analysis}, MGH\cite{biswal2018expert} &Accuracy \\
        & COMET\cite{wang2024contrast} & Multi-level contrast & CNN & Disease detection &AD\cite{escudero2006analysis}, PTB\cite{goldberger2000physiobank}, TDBRAIN\cite{van2022two} &Accuracy, F1, AUROC, AUPRC \\
        & SleepPriorCL\cite{zhang2021sleeppriorcl} &  Expert knowledge incorporation & CNN & Sleep stage classification &SleepEDF\cite{kemp2000analysis}, MASS-SS3\cite{2015Montreal} &Accuracy, F1 \\
        & KDC2\cite{weng2023knowledge} &  Cross-view contrast  & CNN \& GNN & Multiple tasks &SEED\cite{zheng2017identifying}, MMI\cite{schalk2004bci2000}, CHB-MIT\cite{shoeb2009application} &Accuracy \\
        \hline
        \multirow{3}{*}{Mask Autoencoder}    
        & BENDR\cite{kostas2021bendr}  & Temporal-domain mask& CNN \& Transformer & Multiple tasks &MMI\cite{schalk2004bci2000}, BCIC\cite{tangermann2012review}, ERN\cite{margaux2012objective}, SSC\cite{goldberger2000physiobank} &Accuracy \\
        & MAEEG\cite{chien2022maeeg}   & Temporal-domain mask& Transformer & Sleep stage classification &MGH\cite{biswal2018expert} &Accuracy \\
        & Wavelet2vec\cite{peng2023wavelet2vec}    &Frequency-domain mask &ViT& Seizure detection &CHSZ\cite{peng2022tie}, TUSZ\cite{loshchilov2017decoupled} &Accuracy, BCA, F1, MAE \\
    \bottomrule[1pt]
    \end{tabular}
    }
    \label{tab:SSL}
\end{table*}

\section{Discriminative-based EEG Analysis}\label{sec:architecture}
For a more profound comprehension of brain activity, this survey examines advanced architectures, including:
{\textbf{Graph Neural Networks (GNNs)}} in \cref{subsec:gnn}: These networks capitalize on the structural information inherent in brain connectivity to offer deeper insights.
{\textbf{Foundation Models}} in \cref{subsec:foundation_model}: Models pre-trained on extensive datasets and adaptable for specific EEG analysis tasks through fine-tuning.
{\textbf{LLMs-based Methods}} in \cref{subsec:llms}: Leveraging the power of large language models to improve the interpretability of EEG data.

\subsection{Graph Neural Networks}\label{subsec:gnn}
% Introduction to GNN-based Methods
EEG data is a type of multi-channel time series data, in which multiple channels (brain regions) are related to each other, with structural and functional connectivity\cite{BrainNetworks}. Due to brain regions are in non-Euclidean space, graph is the most appropriate data structure to indicate brain connection\cite{jia2020graphsleepnet}. In recent years, graph neural networks(GNN), represented by graph convolutional networks(GCN)\cite{Defferrard2016GCN}, have developed rapidly and become a powerful tool for learning non-Euclidean data representations. They are able to capture intricate relationships inter-variable and inter-temporal, therefore emerging as one of the mainstream frameworks for modeling multivariate time series. Motivated by the success of graph representation learning, a line of studies has utilized GNNs to perform multivariate time series analysis and demonstrate promising results in many downstream tasks such as classification\cite{Wang2023GraphAwareCF}, forecasting\cite{Cai2023MSGNet}, and anomaly detection\cite{Deng2021GDN}. The survey by Jin et al. \cite{Jin2023GNN4TSsurvey} has summarized the application of GNNs in time series analysis, but it does not specifically concentrate on EEG data and only briefly outlines the application in the field of healthcare. In contrast, this paper mainly focuses on EEG data, reviews the recent advances in mainstream EEG analysis tasks with GNNs. It covers a wide range of tasks such as epilepsy detection, sleep staging, and emotion recognition, and sorts out related works from the perspective of EEG graph construction and dependency modeling. All of the methods are summarized in \cref{tab:GNNs}.

\subsubsection{EEG Graph Construction}

% Two methods to construct EEG graphs
In general, each channel in the EEG signal is considered as a node in the graph. Referring to structural connectivity and functional connectivity, the methods for calculating adjacency matrix can be roughly divided into two categories. One is based on the geometry of EEG channels, the other is based on functional connectivity between brain regions. Based on the geometry between the channels, i.e., the anatomical connections between brain regions, previous studies have presented that adjacent brain regions affect each other and the strength of the impact is inversely proportional to the actual physical distance\cite{geometry}. Thus, the adjacency matrix of the graph is constructed from the Euclidean distance between the electrodes, and it is worth noting that this matrix is the same for all EEG clips. The other is based on functional connectivity between brain regions, which captures dynamic brain connections that vary between different EEG clips. It is often calculated based on correlations or dependencies among signals, and the most common methods are Pearson Correlation Coefficient(PCC)\cite{Pearson1903ONTL}, Mutual Information(MI)\cite{Danon2005ComparingCS}, and Phase Locking Value(PLV)\cite{Aydre2013ANO}.

% Utilization of GNNs in EEG Analysis
Tang et al.\cite{Tang2021SelfSupervisedGN} utilizes the above two methods to construct EEGs as graphs and only uses one type of graph as input at a time. Experimental results on the TUSZ v1.5.2 dataset show that the correlation-based graph structure can better localizes focal seizures than the distance-based graph. For a given EEG clip, Ho et al.\cite{Ho2022SelfSupervisedLF} employs four different metrics to construct graphs, including nodes Euclidean distance, randomly connection of nodes, node features correlations, and directed transfer function. The first two are meant to capture the geometry of EEG channels and the last two are for capturing connectivity of brain regions. 

Although the correlation-based graph can be used even when the physical locations of electrodes are unknown, the adjacency matrix is still fixed, which limits its performance to a certain extent. To solve this problem, a lot of research has explored adaptive graph learning strategies. For example, GraphSleepNet\cite{jia2020graphsleepnet} learns the connection relationship between two nodes based on their input features. Specifically, it is implemented through a layer neural network. If the distance between the features of the two nodes is larger, the connection of the two in the adjacency matrix is smaller. And the loss function is defined to be optimized towards this direction. The superiority of adaptive (learnable) adjacency matrix is demonstrated by comparing it with fixed adjacency matrices in the experiment. MSTGCN\cite{Jia2021MSTGCN} uses the adaptive graph learning method proposed by GraphSleepNet\cite{jia2020graphsleepnet}, and also computes the spatial distance-based brain graph. Both views serve as the input of the model to extract features and a concatenate operation is employed to perform feature fusion on the two views. The results of the ablation experiment show that multi-view fusion is more effective than using only one single view. MD-AGCN\cite{Li2021AMA} constructs temporal domain functional brain connectivity and frequency domain functional brain connectivity, respectively. Pearson’s correlation coefficient is used as the connectivity index in the temporal domain. The frequency-domain adjacency matrix is divided into public part and private part. Public part is shared by all of the samples and is set to be trainable parameters, which illustrates the general functional brain connectivity patterns for emotional recognition. Private part is obtained by computing the dot product between two vertexes, and is unique to each sample. Before performing classification, functional brain connections in the two domains are combined together. By visualization of the learned graphs, the results indicate that the model can process global connectivities with the deep layers. BayesEEGNet\cite{Wang2023ABG} considers an electrical impulse between two nodes in the brain as a Poisson process, the countless electrical impulses generated by the brain in a period are represented as an infinite number of connection probability graphs. Then, the countless graphs are coupled into a summary graph by superposition of Poisson distributions, and the summary graph is subsequently transformed into the functional connectivity graph through two three-layer MLPs. By comparing with the adaptive learning strategy proposed by GraphSleepNet\cite{jia2020graphsleepnet}, the connectivity graph obtained in this paper has the best performance in downstream tasks.
\subsubsection{Dependency Modeling and Graph Representation Learning}
Once the EEG graph is constructed, it is often necessary to model the dependencies in the graph to learn the representation that is more discriminative for the downstream task. For example, Tang et al.\cite{Tang2021SelfSupervisedGN} models the spatial dependency in the EEG signals by graph diffusion convolution. And to model the temporal dependency in EEGs, Gated Recurrent Units(GRUs) is employed. Also, in order to learn task-agnostic representations, a self-supervised pretraining method that predicts preprocessed signals for the next time period is proposed. For GraphSleepNet\cite{jia2020graphsleepnet}, a spatial-temporal convolution is designed, which consists of graph convolutions for capturing spatial features and temporal convolutions for capturing temporal context information. Moreover, the attention mechanism is applied in the spatial dimension and the temporal dimension respectively to extract valuable information. BayesEEGNet\cite{Wang2023ABG} also employs the spatial-based graph convolution to aggregate neighbor information directly in the spatial domain. For the emotion recognition task based on multi-modal signals, HetEmotionNet\cite{Jia2021HetEmotionNet} first combines the temporal domain feature vector and the mutual information based adjacency matrix to form a heterogeneous spatial-temporal graph at the current moment, and then stacks the heterogeneous graphs of all time steps to form a heterogeneous graph sequence. Next, the Graph Transformer Network(GTN) is used to model the heterogeneity of multi-modal signals by automatically extracting the meta-paths from the adjacency matrix set. GCN is used to capture the correlation between multi-modal signals, and GRU is applied to extract temporal domain features from the graph sequence obtained after GCN. BrainNet\cite{Chen2022BrainNet} utilizes GCN to model two types of brain wave diffusion processes. Concretely, cross-time diffusion models the propagation of longer epileptic waves between two consecutive time segments. Meanwhile, fast signal spreading within the same time segments of each channel are captured by inner-time diffusion. The experimental results show that both diffusion processes can promote the performance of seizure detection.

There are also methods to mine patterns in a graph by designing self-supervised learning tasks. To capture the correlation patterns in space and time, MBrain\cite{cai2023mbrain} proposes two self-supervised tasks. Instantaneous time shift that is based on multi-channel Contrastive Predictive Coding(CPC) aims to capture the short-term correlations focusing on spatial patterns and delayed time shift is used for temporal patterns in broader time scales. In addition, replace discriminative learning is designed to preserve the unique characteristics of each channel so as to achieve accurate channel-wise seizure prediction. Ho et al.\cite{Ho2022SelfSupervisedLF} leverages a random walk with restart(RWR) technique to create two positive and one negative sub-graphs for every node in every constructed EEG graph, and employs them to perform contrastive learning. Also, a generative learning module is proposed to learn the contextual information hidden in the graph through reconstructing the target node anonymized in the positive sub-graphs, using the other node features and edges of the sub-graph. To promote spatial consistency in multiple sensors, GCC\cite{Wang2023GraphAwareCF} proposes novel graph augmentations including node augmentations and edge augmentations, to augment sensors and their correlations respectively. Next, a graph contrasting method is designed. Node-level Contrasting is achieved by contrasting sensors in different views within each sample while Graph-level Contrasting is achieved by contrasting the samples within each training batch. Through these two contrasting procedures, robust sensor-level features and global-level features can be learned.

\begin{table*} [t]
   \caption{ Summary of representative GNN-based Methods for EEG Analysis.}
   \label{tab:GNNs}
    \centering
    \begin{adjustbox}{width=2\columnwidth,center}
    \begin{tabular}{lccccccc}
    \toprule[1pt]
        \textbf{Task} & \textbf{Method} & \textbf{Graph Construction} & \textbf{Spatial Module} & \textbf{Temporal Module} & \textbf{Training} & \textbf{Datasets} & \textbf{Metric}\\ 
        \midrule
        \multirow{2}{*}{Sleep Stage Classification} 
        & GraphSleepNet\cite{jia2020graphsleepnet}  & Learned & Spectral, Attention & CNN, Attention & - & MASS-SS3\cite{2015Montreal} & Accuracy, F1-score, Kappa\\ 
        & MSTGCN \cite{Jia2021MSTGCN} & Learned & Spectral, Attention & CNN, Attention & - & ISRUC-S3\cite{KHALIGHI2016180}, MASS-SS3\cite{2015Montreal} & Accuracy, F1-score, Kappa\\ 
        \hline
        \multirow{2}{*}{Emotion Recognition}& HetEmotionNet \cite{Jia2021HetEmotionNet} & FC & Spectral & GRU & - & DEAP\cite{koelstra2011deap}, MAHNOB-HCI\cite{MAHNOBHCI2012} & Valence, Arousal\\ 
        & MD-AGCN \cite{Li2021AMA}  & FC, Learned & Spatial & - & - & SEED, SEED-IV, SEED-V\cite{zheng2017identifying} & Accuracy\\ 
        \hline
        \multirow{4}{*}{Seizure Detection} &Tang et al. \cite{Tang2021SelfSupervisedGN}& SC, FC & Spatial, Spectral & GRU & Generative Learning & TUSZ\cite{Obeid2016TheTU} & AUROC, F1-score\\ 
        &BrainNet \cite{Chen2022BrainNet} & Learned & Spatial & - & Contrastive Learning & Private data & Precision, Recall, F1-score, F2-score, AUC\\ 
        &MBrain \cite{cai2023mbrain} & Learned & - & - & Contrastive Learning & Private data, TUSZ\cite{Obeid2016TheTU} & Precision, Recall, F1-score, F2-score\\ 
        &EEG-CGS \cite{Ho2022SelfSupervisedLF}  & SC, FC & - & - & Contrastive and Generative Learning & TUSZ\cite{Obeid2016TheTU} & ROC-AUC, Precision, F1-score, Sensitivity, Specificity\\ 
        \hline
         Sleep Stage Classification&\multirow{2}{*}{BayesEEGNet \cite{Wang2023ABG}} & \multirow{2}{*}{Learned} & \multirow{2}{*}{Spatial} & \multirow{2}{*}{-} & \multirow{2}{*}{-} & \multirow{2}{*}{MASS-SS3\cite{2015Montreal}, ISRUC-S3\cite{KHALIGHI2016180}, SEED\cite{zheng2017identifying}} & \multirow{2}{*}{Accuracy, F1-score, Kappa}\\
        and Emotion Recognition&&&& \\
    \bottomrule[1pt]
    \end{tabular}
    \end{adjustbox}
        \caption*{\small Graph Construction: "SC" and "FC" denote "structural connectivity" and "functional connectivity", respectively. "Learned" indicates that the graph structure is learned from data.}
\end{table*}
\subsection{Foundation Models}\label{subsec:foundation_model}
% \section{Foundation Model} 
Foundation models (FMs)\cite{bommasani2021opportunities}, often known as large-scale pretrained models, are advanced neural networks trained on extensive datasets. These models possess a vast range of general knowledge and can recognize numerous patterns. As a result, they offer a flexible and comprehensive foundation for addressing various tasks across multiple domains. ChatGPT\cite{brown2020language} is the most famous textural foundation model that has a powerful ability to understand and generate natural language texts, and can perform a variety of natural language processing tasks, including text classification, sentiment analysis, machine translation, etc., showing extremely high flexibility and generalization capabilities. CLIP \cite{radford2021learning} and SAM \cite{kirillov2023segment} are representative visual foundation models, which exhibit robust general understanding and reasoning performance. Foundation models consistently demonstrate high performance in diverse domains, from natural language processing to computer vision, showcasing their versatility and the potential to revolutionize the way AI systems interact with and understand the world.

In the field of EEG data processing, researchers usually proposed specially designed methods or models for specific data or tasks. However, data annotation in the medical field is more difficult and expensive than in other fields. As a result, the size of EEG medical data sets is usually small, which greatly restricts the capabilities of the model \cite{Wagh2020EEGGCNNAE,Chen2022BrainNetEW}. The emergence of large language models provides a new solution for the processing of biological signal data such as EEG. Recently, a lot of work has begun to draw on the ideas of large language models, using a large amount of unlabeled data and unsupervised pre-training methods to build foundation models for EEG or biological signal data \cite{Zhang2023BrantFM,Cui2023NeuroGPTTA,Abbaspourazad2023LargescaleTO,Jiang2024LargeBM,zhang2024brant,yuan2024brant, chen2024eegformer}. These foundation models have learned a lot of knowledge about time series signals, can well represent EEG data, have generalization capabilities that previous models did not have, and can achieve excellent performance on different downstream tasks. 
% Below, we will outline the existing work related to foundation models in the field of EEG signals from the three important elements of data (which are presented in \cref{sec:dataset_and_metric}), model structure, and training methods. 
Below, we outline the existing work related to foundation models in the field of EEG signals, considering the three important elements: data, model structure, and training methods. While the datasets themselves are thoroughly described in \cref{sec:dataset_and_metric}, this chapter will focus on how they are used in the process of EEG foundation models established. 

While the datasets are crucial and will be extensively discussed, this chapter is dedicated to the presentation of the models and training methodologies. The summary of existing foundation models is shown as \cref{tab:foundation_models}.

\begin{table*}
    \centering
    \caption{Summary of Foundation models for EEG Analysis}
    \tiny
    \label{tab:foundation_models}
    \begin{tabular}{ccccc}
    \toprule[1pt]
        \textbf{Method}  & \textbf{Model Structure}  &  \textbf{Training} & \textbf{Datasets} & \textbf{Metric}\\
        \midrule
         BrainBERT \cite{Wang2023BrainBERTSR} & Transformer blocks & Masked Autoencoder & Private data  & AUC\\
         Neuro-GPT \cite{Cui2023NeuroGPTTA} & Convolutional blocks + Transformer blocks  &  Future Forecast & TUH EEG corpus\cite{Obeid2016TheTU} & MSE, Accuracy\\
         Brant \cite{Zhang2023BrantFM} & Transformer blocks & Masked Autoencoder & Private data & MSE, MAE, F1, F2 \\
         BFM \cite{Abbaspourazad2023LargescaleTO} & Convolutional blocks & Contrastive Learning & AHMS corpus\cite{2019AppleH} & AUC, MAE\\
         LaBraMs \cite{Jiang2024LargeBM}  & Convolutional blocks + Transformer blocks & Masked Autoencoder & Public data + Private data & Accuracy, AUROC, F1\\
    \bottomrule[1pt]
    \end{tabular}
\end{table*}

\subsubsection{Model Structure}
With the rapid development of deep learning, many model structures have emerged, such as Convolutional Neural Network (CNN) \cite{LeCun1998GradientbasedLA}, Recurrent Neural Network (RNN) \cite{Zaremba2014RecurrentNN}, Transformers\cite{Vaswani2017AttentionIA}, Mamba \cite{Gu2023MambaLS}, etc. How to design a model structure suitable for processing time series signals is the top priority in building a foundation model. A good structure can allow the foundation model to better understand and learn the information and knowledge in time series signals. Most of the existing EEG foundation models construct the main model by stacking Transformer layers or convolutional blocks. Because both structures have strong scalability and are suitable for mining information in time series signals.

Brant \cite{Zhang2023BrantFM} has two encoders, temporal encoder and spatial encoder. The temporal encoder contains a 12-layer Transformer encoder and the spatial encoder contains a 5-layer Transformer encoder. They are used to capture the time correlation and channel correlation in time series signals, respectively. Salar et al. \cite{Abbaspourazad2023LargescaleTO} built the foundation model based on an EfficientNet-style 1D convolutional neural network. Neuro-GPT \cite{Cui2023NeuroGPTTA} and LaBraM \cite{Jiang2024LargeBM} use both convolutional layers and Transformers layers. They first use a small number of convolutional layers to preliminarily extract the features of time series signals and transform their dimensions, and then use a large number of Transformers layers to further capture the correlation between different sequence patches and better represent time series signals.

Since the input of the Transformer layer is tokens, and the time series data is a continuous value, the foundation model needs to convert the time series data into patches before subsequent calculations can be performed. A common approach is to split the original data by a fixed window size and a fixed strides. Specifically, given a neural signal $\boldsymbol{x} \in \mathbb{R}^{N \times C}$, where $N$ is the number of timestamps and $C$ is the number of electrode channels, we divide $x$ with window size $M$ and stride $S$ to generate a set of patches $\boldsymbol{p} \in \mathbb{R}^{N_p \times C \times M}$, where $N_{p} \equiv\left\lfloor\frac{N-M}{S}\right\rfloor$ is the number of patches in each channel. After obtaining the segmented patches, additional position or frequency encoding information is usually added to them to help the model learn better. Some researchers \cite{Jiang2024LargeBM} also map each patch to a fixed codebook in order to make the foundation model have a fixed vocabulary like a large language model. Specifically, it first represents the patch and then utilizes quantizer to quantize all the patch representations into the neural codebook embeddings. The codebook looks up the nearest neighbor of each patch in the neural codebook.

The parameter size of the existing foundation models in the EEG field is usually between tens and hundreds of millions, which is still relatively small compared to the parameters of large language models. This may be because the amount of EEG data is still much smaller than text data. However, we believe that with the continuous development of the field, the scale of the foundation model will continue to increase, and its capabilities will continue to increase.
\subsubsection{Training Methods}
In order for the model to learn useful knowledge from massive amounts of unlabeled data, it is essential to design an effective training method. A good training method is like a good teacher, which can make the learning process more efficient.

Existing foundation models are all pre-trained using self-supervised methods. One of the mainstream approaches is to use masked autoencoder (MAE) as a pre-training task \cite{Zhang2023BrantFM,Jiang2024LargeBM,Wang2023BrainBERTSR}. MAE has been proven to be a simple and effective method in many fields, which trains model to reconstruct the whole input given its partial observation. In this way, the foundation model can be forced to infer the whole from partial information, so that the model can learn powerful representation capabilities.

There is another pre-training method that is similar to MAE, which can be understood as masking only the latter part of the input. During the training process, the model predicts the future situation based on the historical content of the time series data \cite{Cui2023NeuroGPTTA}. Its goal is actually the same as the short-term or long-term prediction in the downstream task. Therefore, the foundation model pre-trained by this method usually has strong predictive ability, which can capture regularities from historical time series data.

Another type of work uses contrastive learning to train the foundation model. The core idea is to learn how to effectively distinguish similar (positive) and dissimilar (negative) data points by comparing data samples, so as to optimize the data representation or feature vector. This method can help the model capture the intrinsic structure and relationship between data, thereby improving its generalization ability on downstream tasks. For example, Salar et al. \cite{Abbaspourazad2023LargescaleTO} constructed positive and negative pairs at the participant level. Specifically, the positive pairs are selected as augmented views of two different segments from the same participant, while the segments from different subjects are regarded as negative samples. Through this training method, the model can not only acquire strong representation capabilities, but also enhance its generalization ability on different subjects.

Using various pre-training methods, the foundation model can acquire enough knowledge from a large amount of unlabeled data. Therefore, it only needs to be fine-tuned with a small amount of data to be well adapted to various downstream tasks. It can even have zero-shot capabilities like a large language model. This makes it possible to build a universal EEG foundation model.
\subsection{LLMs-based Methods}\label{subsec:llms}
% Introduction to LLMs-based Methods
Large Language Models (LLMs)\cite{touvron2023llama,touvron2023llama2,OpenAI2023GPT4TR} have revolutionized the field of natural language processing (NLP) by demonstrating remarkable capabilities in understanding, generating, and translating human language. The application of LLMs in EEG analysis represents a novel and innovative approach to interpreting complex brain signals. Unlike traditional machine learning methods, LLMs can be fine-tuned with relatively small amounts of task-specific data, making them particularly well-suited for the analysis of EEG data, which can be challenging to annotate and label.

    % [Unimodal-based\\ Models
    %                         [Victor\cite{iapascurta2023nlp} PromptCast\cite{xue2023promptcast} TEMPO\cite{cao2023tempo} \\LLM4TS\cite{chang2023llm4ts} Time-LLM\cite{jin2023time} $S^2$IP-LLM\cite{pan2024textbf} \\GPT4TS\cite{zhou2023one} TEST\cite{sun2023test} Zhang\cite{zhang2024word}, leaf, text width=25em]
    %                     ]
    %                     [Multimodal-based\\ Models
    %                         [EEG-To-Text \cite{wang2022open} MTAM\cite{qiu2023can} METS\cite{li2024frozen}  \\GPT4MTS\cite{jia2024gpt4mts} ESI\cite{yu2024ecg} InstructTime\cite{cheng2024advancing} \\CrossTimeNet\cite{cheng2024learning}  CALF\cite{liu2024taming} EEG-GPT\cite{kim2024eeg}  K-Link\cite{wang2024k} , leaf, text width=25em]
    %                     ]
    %                 ]
\begin{table*}
    \centering
   \begin{adjustwidth}{0.3cm}{2.5cm}
    \caption{Summary of LLMs-based Methods for EEG Analysis}
    \label{tab:llm}
    \tiny
     \scalebox{0.8}{
    \begin{tabular}{ccccccc}
       \toprule[1pt]
       \textbf{Modality} & \textbf{Method}  & \textbf{Task} & \textbf{Language model} & \textbf{Training} & \textbf{Datasets} & \textbf{Metric}\\
     \hline
        % \multirow{9}{*}{Unimodal} & EEG-To-Text \cite{wang2022open}  & Sentiment Classification & BART\cite{lewis2019bart} & Reconstruction cross-entropy & & \\
          \multirow{8}{*}{\textbf{Uni}} & Victor\cite{iapascurta2023nlp}   & Prediction & BERT\cite{Vaswani2017AttentionIA}& Cross-entropy &American Epilepsy Society\cite{14AmericanEpilepsySociety} & {AUC, Accuracy}\\
        \cline{2-7} 
                & PromptCast\cite{xue2023promptcast}    & \multirow{5}{*}{Forecast}  & T5\cite{raffel2020exploring}, BART\cite{lewis2019bart}, etc. & Template-Based Prompting & PISA& \multirow{5}{*}{MAE, MSE} \\
        \cline{4-6} 
                & TEMPO\cite{cao2023tempo}    &  & GPT2\cite{radford2019language} , T5\cite{raffel2020exploring}, LLaMA\cite{touvron2023llama2}, etc. & STL\cite{cleveland1990stl}, LoRA\cite{hu2021lora} & \multirow{4}{*}{\textbf{LTSF}\cite{Wu2021AutoformerDT}}
 & \\
                & LLM4TS\cite{chang2023llm4ts}  &  & GPT2\cite{radford2019language} & Autoregressive & & \\
                & Time-LLM\cite{jin2023time}  &  & LLaMA\cite{touvron2023llama2} & Reprogramming, Prompt-as-Prefix & & \\
                & $S^2$IP-LLM\cite{pan2024textbf} &  & GPT2\cite{radford2019language}  & Partial fine-tune & & \\
        \cline{2-7} 
                & GPT4TS\cite{zhou2023one} &Classification, Forecast, etc.& GPT2\cite{radford2019language}  & Partial fine-tune &\multirow{2}{*}{UEA\cite{Bagnall2018TheUM}} & \multirow{2}{*}{Accuracy} \\
                & TEST\cite{sun2023test} &Classification & GPT2\cite{radford2019language}  & Contrastive learning & & \\
        \cline{2-7}
                & Zhang\cite{zhang2024word} & Eye-tracking & GPT-3.5’s and GPT-4’s APIs  &  LLM agent &ZuCo\cite{hollenstein2018zuco} &{AUC, Accuracy}\\
                \hline
        \multirow{10}{*}{\textbf{Mul}} 
         & EEG-To-Text \cite{wang2022open}  & Sentiment Classification & BART\cite{lewis2019bart} & Reconstruction, cross-entropy &ZuCo\cite{hollenstein2018zuco} &{AUC, Accuracy}\\
         \cline{2-7}
                & MTAM\cite{qiu2023can}  &  Analysis, Relation Detection & Transformer&  CCA, WD &{K-EmoCon\cite{Park2020KEmoConAM}, ZuCo\cite{hollenstein2018zuco}}&Precision, Recall,F1-score, Accuracy\\
                & METS\cite{li2024frozen} & Clinical Diagnosis & ClinicalBert\cite{alsentzer2019publicly} & Contrastive Learning & PTB-XL\cite{Wagner2020PTBXLAL},MIT-BIH\cite{Moody2001TheIO}& Precision, Recall,F1-score, Accuracy\\
                & GPT4MTS\cite{jia2024gpt4mts}    & Forecast &BERT\cite{Vaswani2017AttentionIA}, GPT2\cite{radford2019language}& Partial fine-tune & GDELT& MAE, MSE\\
                & ESI\cite{yu2024ecg}     & Diagnosis &BioLinkBERT\cite{yasunaga2022linkbert}& Contrastive Learning, RAG & {CSX\cite{Zheng2020OptimalMA}, PTB-XL\cite{Wagner2020PTBXLAL}, MIT-BIH\cite{Moody2001TheIO}}&AUC, Accuracy\\
                & InstructTime\cite{cheng2024advancing}   & Classification& GPT2\cite{radford2019language} &VQ-VAE\cite{Oord2016WaveNetAG}, Full fine-tuning &EEG\cite{Andrzejak2001IndicationsON}, ECG\cite{Cheng2023TimeMAESR}, HAR\cite{Anguita2013APD}, FD\cite{Liu2021GatedTN} & F1-score, Accuracy\\
                & CrossTimeNet\cite{cheng2024learning}   &Classification &BERT\cite{Vaswani2017AttentionIA}& VQ-VAE\cite{Oord2016WaveNetAG} & EEG\cite{Andrzejak2001IndicationsON}, ECG\cite{Cheng2023TimeMAESR}, HAR\cite{Anguita2013APD}&  F1-score, Accuracy \\
                & CALF\cite{liu2024taming}    &Forecast &GPT2\cite{radford2019language} & Distill Knowledge &{\textbf{LTSF}\cite{Wu2021AutoformerDT}} & MAE, MSE\\
                & EEG-GPT\cite{kim2024eeg}    &Classification &Vinci GPT-3&  prompt-completion API & TUH EEG Corpus\cite{Obeid2016TheTU}& AUC-ROC\\
                & K-Link\cite{wang2024k}   & Forecast&CLIP-Text\cite{radford2021learning}&Contrasting Learning& {\textbf{LTSF}\cite{Wu2021AutoformerDT}} &  MAE, MSE\\
    \bottomrule[1pt]
    \end{tabular}
    }
    \end{adjustwidth}
    \caption*{\small \textbf{Uni} represents unimodal-based methods, and \textbf{Mul} represents multimodal-based methods. \\ \textbf{LTSF} contains ETTh1/h2/m1/m2, Weather, Electricity, Traffic}
\end{table*}
% Utilization of LLMs in EEG Analysis
The integration of LLMs into EEG analysis can take two forms:
\textbf{Unimodal-based Models}: These approaches use LLMs as feature extractors for EEG data sets, which are of a single modality, implicitly leveraging the semantic knowledge that these models contain. Here, LLMs can be fine-tuned to classify different neurological states or forecast outcomes based on EEG data with Parameter Efficient Fine-Tuning (PEFT) techniques\cite{han2024parameter}, such as LoRA\cite{hu2021lora} or soft prompt\cite{lester2021power}. Their proficiency in handling sequential data makes them particularly adept at time-series analysis.
\textbf{Multimodal-based Models}: These approaches deals with multi-modal data, where EEG is paired with text using LLMs through knowledge distillation\cite{hinton2015distilling} or cross-modal contrastive learning\cite{radford2021learning}.
% Recently, there has been significant progress in adapting LLMs for general time series analysis. As we all know, EEG is fundamentally time series data. So we believe that advancements in general time-series can be successfully applied to EEG data analysis in the near future. Thus, we will also provide a brief overview of some mainstream methods in general time series analysis.
What's more, there has been significant progress in adapting LLMs for general time series analysis\cite{jiang2024empowering,zhang2024large,liang2024foundation}. For those familiar with the field, it is well understood that EEG data is a type of time series data. Given this, we are confident that the advancements made in general time series analysis can be successfully applied to EEG data analysis in the near future. Consequently, we intend to provide a brief overview of some mainstream methods currently utilized in general time series analysis.  All of the methods are summarized in \cref{tab:llm}.
 % \textbf{Interpretation and Summarization}: LLMs can be trained to interpret EEG patterns and generate natural language summaries of brain activity. This can aid clinicians and researchers in understanding complex EEG readings more intuitively.
\subsubsection{Unimodal-based Models}These approaches use LLMs as the backbone, harnessing the models' inherent semantic understanding. Some works adapts them for time-series forecasting tasks. Victor et.al\cite{iapascurta2023nlp} first employs the Kolmogorov-Chaitin algorithm to convert EEG data into a text-like format, and then constructs a machine-learning model based on language models to predict epilepsy. PromptCast\cite{xue2023promptcast} introduces an innovative "codeless" approach to time series forecasting, offering a fresh perspective that moves away from the sole emphasis on creating complex architectures. TEMPO\cite{cao2023tempo} concentrats exclusively on time series forecasting while integrating additional intricate elements such as time series decomposition and soft prompts. LLM4TS\cite{chang2023llm4ts} proposes a two-stage fine-tuning framework for time-series forecasting, addresses challenges in incorporating LLMs with time-series data. Time-LLM\cite{jin2023time} reprograms time series by incorporating the source data modality and utilizing natural language-based prompting, which unlocks the potential of LLMs as efficient time series machines. $S^2$IP-LLM\cite{pan2024textbf} leverages LLMs by aligning their semantic space with time series embeddings to enhance time series forecasting through semantic space-informed prompt learning. The vast majority of existing research in the field has been centered on time-series forecasting tasks. This focus may stem from the inherent similarities between the autoregressive processes of LLMs and the forecasting nature of time-series prediction models. In other words, the resemblance lies in the fact that both types of models rely on historical data (or context) to make predictions about future data points (or words in the case of LLMs). In addition to forecasting, a few works have adapted LLMs for time-series classification. GPT4TS\cite{zhou2023one} presents a unified framework with frozening the self-attention and feedforward layers of the residual blocks in the LLMs and fine-tuning the layer norm layer. TEST\cite{sun2023test} converts time-series data into a format suitable for pre-trained LLMs by employing a three-level contrast approach, which includes instance-wise, feature-wise, and text-prototype-aligned contrasts. Zhang et al.\cite{zhang2024word} utilize LLMs to generate labels that guide a new reading embedding representation for EEG, enabling the prediction of human reading comprehension at the word level. In summary, recent studies reflect a burgeoning interest in harnessing the capabilities of LLMs for time-series analysis by integrating them into the architecture in ways that capitalize on the inherent strengths of LLMs. 

\subsubsection{Multimodal-based Models}
% multimodal
% In addition to methods that concentrate exclusively on time series data, there have been concerted efforts to create multi-modal application. METS\cite{li2024frozen} use a trainable ECG encoder and a frozen language model to embed paired ECG and automatically machine generated clinical reports separately by multimodal contrastive learning. GPT4MTS\cite{jia2024gpt4mts} proposes a multimodal time series dataset for news impact forecasting and introduces a prompt-based large language model (LLM) framework to utilize both numerical values and textual information. ESI\cite{yu2024ecg}  integrates a retrieval-augmented generation (RAG) pipeline to get external medical
% knowledge to enrich textual descriptions. InstructTime\cite{cheng2024advancing} formulate the classification of time series as a multimodal understanding task, in which both task-specific instructions and raw time series are treated as multimodal inputs while the label information is represented by texts. CrossTimeNet\cite{cheng2024learning} designs a time series tokenization module that effectively convert the raw time series into a sequence of discrete tokens based on a reconstruction optimization process. CALF\cite{liu2024taming} develops a cross-modal match module to first align cross-modal input distributions  between textual and temporal data and further align modality distribution gap in both feature and output spaces.
In addition to methods that focus solely on time series data, there have been significant efforts to develop multi-modal applications. EEG-To-Text \cite{wang2022open} presents a novel framework using LLMs to extend brain-to-text decoding to open vocabulary and achieve zero-shot sentiment classification. MTAM\cite{qiu2023can} uses a multimodal transformer alignment model to investigate the correlation between EEG data and language, enabling the observation of synchronized representations across these modalities and utilizing these aligned representations for various downstream tasks. METS\cite{li2024frozen} employs a trainable ECG encoder alongside a frozen language model to embed paired ECG signals and automatically generated clinical reports separately through multimodal contrastive learning. GPT4MTS\cite{jia2024gpt4mts} introduces a multimodal time series dataset for news impact forecasting and proposes a prompt-based LLM framework that leverages both numerical values and textual information. ESI\cite{yu2024ecg} integrates a retrieval-augmented generation (RAG) pipeline to obtain external medical knowledge, thereby enriching textual descriptions. InstructTime\cite{cheng2024advancing} formulates the classification of time series as a multimodal understanding task, treating both task-specific instructions and raw time series data as multimodal inputs, with label information represented in text form. CrossTimeNet\cite{cheng2024learning} designs a time series tokenization module that effectively converts raw time series data into a sequence of discrete tokens based on a reconstruction optimization process. CALF\cite{liu2024taming} develops a cross-modal match module to align cross-modal input distributions between textual and temporal data, further bridging the modality distribution gap in both feature and output spaces. EEG-GPT\cite{kim2024eeg} offers intermediate reasoning steps and coordinate EEG tools across different scales, providing a transparent, interpretable, step-by-step analysis that enhances trustworthiness in clinical application. K-Link\cite{wang2024k} proposes a framework that enriches a signal-derived graph by integrating a knowledge-link graph, which is constructed using LLMs, through the process of graph alignment. In summary, these efforts underscore the potential of integrating time series methods with the capabilities of LLMs to develop more robust and informative models. This is achieved through techniques that utilize a dual-tower architecture, such as cross-modal contrastive learning and knowledge distillation processes.

\section{Generative-based EEG Analysis}\label{sec:generative}
In this section, we will delve into innovative generative applications that utilize EEG data to produce images or text, providing novel approaches to the visualization and understanding of brain activity.In this section, we explore the performance of EEG analysis methods on multi-modal generation tasks. Previous works have proved that EEG signal contain abundant semantics. It’s intuitively that we can reconstruct the semantics information from EEG signal instead of just catch their representation from raw data with the help of generative model such as GANs\cite{goodfellow2020generative}, Diffusion Models\cite{ho2020denoising} and Transformers based models. All of the methods are presented in \cref{tab:eeg_generation}.

% \section{Generation}
% In this section, we explore the performance of EEG analysis methods on multi-modal generation tasks. Previous works have proved that EEG signal contain abundant semantics. It’s intuitively that we can reconstruct the semantics information from EEG signal instead of just catch their representation from raw data with the help of generative model such as GANs\cite{goodfellow2020generative}, Diffusion Models\cite{ho2020denoising} and Transformers based models.
\begin{table*}[htb]
    \centering
    \caption{Summary of EEG-To-Modality Generation Models}
    \resizebox{2\columnwidth}{!}{
    \begin{tabular}{ccccccc}
    \toprule[1pt]
       \textbf{Modality} & \textbf{Method}  & \textbf{Encoder} & \textbf{Decoder} & \textbf{Pretrained} & \textbf{Dataset} &\textbf{Eval Metric} \\
       \hline
        \multirow{6}{*}{Image}   & Brain2Image\cite{kavasidis2017brain2image}  & LSTM & VAE & Classification & Spampinato\cite{spampinato2017deep} & IS\\
                & ThoughtViz\cite{tirupattur2018thoughtviz}   & CNN  & GAN & Classification & Kumar\cite{kumar2018envisioned} & IS \& Accuracy\\
                & EEG2Image\cite{singh2023eeg2image}    & LSTM & DCGAN & Constrastive learning & Kumar\cite{kumar2018envisioned} & IS \\
                & EEGStyleGAN-ADA\cite{singh2024learning}  & LSTM & SyleGAN-ADA & Constrastive learning& Spampinato\cite{spampinato2017deep} Kumar\cite{kumar2018envisioned}  Kaneshiro\cite{kaneshiro2015representational} & IS \& FID \& KID\\
                & DreamDiffusion\cite{bai2023dreamdiffusion}  & VQ & LDM & MAE & Spampinato\cite{spampinato2017deep} & Accuracy\\
                & NeuroImagen\cite{lan2023seeing} &Saliency Map, BLIP & LDM & Map & Spampinato\cite{spampinato2017deep} & IS \& Accuracy \& SSIM\\
                \hline
        \multirow{4}{*}{Text}    & EEG-To-Text\cite{wang2022open}  &Transformer &\multirow{4}{*}{BART\cite{lewis2019bart}} & Map & ZuCo\cite{hollenstein2018zuco} &\multirow{4}{*}{BLEU-N \& ROUGE-1} \\
                & EEG2Text\cite{liu2024eeg2text}  & Convolutional Transformer& & MAE & ZuCo\cite{hollenstein2018zuco} Image-EEG\cite{gifford2022large}\\
                & E2T-PTR\cite{wang2024enhancing}   & Multi-stream Transformer& & MAE& ZuCo\cite{hollenstein2018zuco} \\
                & DeWave\cite{duan2024dewave}    &VQ-VAE && -& ZuCo\cite{hollenstein2018zuco} \\
                \hline
        \multirow{2}{*}{Others}  & ETCAS\cite{guo2023end}    & - &Dual-DualGAN & -&Privated data& Accuracy \& PCC \& MCD\\
                & NDMusic\cite{daly2023neural}    & - &BiLSTM& - &MusicAffect &Rank accuracy \\
    \bottomrule[1pt]
    \end{tabular}
    }
    \label{tab:eeg_generation}
\end{table*}
\subsection{Image Generation}

\begin{table}[h]
    \centering
    \caption{EEG-Image Dataset for Image Generation}
    \tiny
    \label{tab:eeg_dataset}
    \begin{tabular}{cccc}
    \toprule[1pt]
    \multirow{2}{*}{\diagbox{\textbf{Item}}{\textbf{Dataset}}}
    & \textbf{Spampinato}& \textbf{Kumar }& \textbf{Kaneshiro } \\
    & \cite{spampinato2017deep}& \cite{kumar2018envisioned}& \cite{kaneshiro2015representational} \\
    \hline
    Classes & 40 & 30 & 6 \\
    Subjects & 6 & 23 & 10 \\
    Channels & 128 & 14 & 128 \\
    Quantity & 2000 & 30 & 72 \\
    Frequency (Hz) & 1000 & 2048 & 1000 \\
    Time(s) & 0.5 & 10 & 0.5 \\
    Pause(s) & 10 & 20 & 0.75 \\
    \bottomrule[1pt]
    \end{tabular}
\end{table}

% \begin{table}[h]
% \centering
% \caption{The average performance of the domain generalization experiment on SEEG datasets.}
% \begin{tabular}{lc}
% \toprule
% Model Variant & Accuracy (\%) \\
% \midrule
% VanillaNet       & 54.86 \\
% VanillaNet + inter-modal loss  & 63.37 \\
% VanillaNet + intra-modal loss  & 64.58 \\
% VanillaNet + CrossEpilepsy (ours) & 69.23 \\
% \hline
% Efficientnet-B0~\cite{tan2019efficientnet} + CrossEpilepsy & 63.74\\
% Vit-Tiny~\cite{dosovitskiy2020image} + CrossEpilepsy & 65.46 \\
% \hline
% \bottomrule
% \end{tabular}
% \label{table:exp2}
% \end{table}
EEG-Image generation tasks typically follow the Map-Train-Finetune paradigm, which ensures high semantic fidelity but poses challenges in training and fine-tuning. Brain2Image\cite{kavasidis2017brain2image} addresses these challenges by dividing the EEG-Image generation task into two distinct phases. In the first phase, Brain2Image encodes EEG signals into a lower-dimensional feature vector for conditioning in image generation. Specifically, a standard LSTM layer followed by a nonlinear layer is trained to classify the EEG signals, serving as the encoder. An additional fully-connected layer is then added to ensure the learned EEG feature vector follows a Gaussian distribution, as required by Variational Autoencoders (VAEs). In the second phase, for each EEG sequence provided to the encoder, Brain2Image uses the encoder's output to train the VAE's decoder to generate images corresponding to what the subject is observing at that precise moment.
Compared to Brain2Image, ThoughtViz\cite{tirupattur2018thoughtviz} employs a 1D-CNN followed by a 2D-CNN for EEG classification as an encoder. Building on the traditional GAN architecture, ThoughtViz introduces a pre-trained classifier to classify the samples generated by the generator. The generator loss in ThoughtViz incorporates both the discriminative loss from the discriminator and the classification loss from the classifier.

Unlike training the EEG encoder through a supervised classification task, EEG2Image\cite{singh2023eeg2image} and EEGStyleGAN-ADA\cite{singh2024learning} employ a triplet loss-based contrastive learning approach in their proposed frameworks for EEG feature learning. The triplet loss function aims to minimize the distance between data points with the same labels while maximizing the distance between data points with different labels. This approach prevents the EEG encoder from compressing the representations into small, indistinct clusters. EEG2Image utilizes a Conditional DCGAN\cite{radford2015unsupervised} architecture with hinge loss for stable training, whereas EEGStyleGAN-ADA employs StyleGAN-ADA\cite{karras2020training} with adaptive discriminator augmentation. This augmentation helps the discriminator effectively learn from limited data by augmenting real images during training.

With the powerful generative capabilities of Diffusion Models, an increasing number of researchers are applying these models to the EEG-Image generation task. DreamDiffusion\cite{bai2023dreamdiffusion}, for instance, collects a large-scale unlabeled EEG dataset from the MOABB\cite{jayaram2018moabb} platform and uses the MAE method for brain pretraining. During the fine-tuning stage, DreamDiffusion employs a projection layer to align brain latent representations with CLIP-Image semantic information. NeuroImagen\cite{lan2023seeing}, on the other hand, uses detail and semantic extractors to map EEG signals to pixel and CLIP-Text priors, which are then decoded by a pretrained Stable Diffusion model following the image-to-image pipeline.

\subsection{Text Generation}
% Unveiling Thoughts: A Review of Advancements in EEG Brain Signal Decoding into Text
Unlike EEG-image generation, EEG-text generation is a sequence-to-sequence process. Inspired by machine translation applications using pretrained BART\cite{lewis2019bart}, Wang et al.\cite{wang2022open} consider the human brain as a unique type of encoder. They treat each EEG feature sequence as an encoded sentence by the human brain and then train an additional encoder to map the brain's embeddings to the embeddings from the pretrained BART model.
Instead of using the word-level EEG features crafted based on the eye-tracking data like \cite{wang2022open}, EEG2Text\cite{liu2024eeg2text} directly use the sentence-level EEG signals as input to the model. Specifically, EEG2Text leverages EEG pre-training to enhance the learning of semantics from EEG signals and proposes a multiview transformer to model the EEG signal processing by different spatial regions of the brain.

Wang et al.\cite{wang2024enhancing} introduced CET-MAE, a model that combines contrastive learning and masked signal modeling via a multi-stream encoder. It effectively learns EEG and text representations by balancing self-reconstructed latent embeddings with aligned text and EEG features. They also propose an EEG-to-Text decoding framework using Pretrained Transferable Representations, leveraging LLMs for language understanding and generation, and fully utilizing the pre-trained representations from CET-MAE.
To address significant distribution variances in EEG waves across individuals and rectify order mismatches between raw wave sequences and text, DeWave\cite{duan2024dewave} uses a vector quantized variational encoder. This encoder transforms EEG waves into a discrete codex, linking them to tokens based on proximity to codex book entries. DeWave is the first to introduce discrete encoding into EEG signal representation, benefiting both word-level EEG features and raw EEG wave translation.

\subsection{Others}
In addition to image and text generation, many other EEG-to-modality generation tasks deserve attention. ETCAS\cite{guo2023end}, an end-to-end GAN model tailored for EEG-based sound generation tasks, introduces a Dual-DualGAN to directly map EEG signals to speech signals. NDMusic\cite{daly2023neural} adopts an end-toend bidirectional LSTM (BiLSTM) architecture to establish a direct mapping from fMRI-informed EEG signals to music signals.
\section{Datasets and Metrics}\label{sec:dataset_and_metric}
The analysis of spatio-temporal EEG data relies heavily on the availability of high-quality datasets and robust evaluation metrics. This section provides an overview of the most widely used datasets and the key metrics employed to assess the performance of various EEG analysis models.

\subsection{Datasets}
% reference 
% https://paperswithcode.com/datasets?q=eeg&v=lst&o=match&mod=eeg&page=1
%  用于什么场景，多少个被试，多少数据
\subsubsection{\textbf{Publicly Available EEG Datasets}}
Several publicly available EEG datasets have been instrumental in advancing the field. These datasets vary in their focus, including different cognitive tasks, subject demographics, and recording conditions. \\
\textbf{Discriminative EEG Task Dataset}: These datasets are typically employed for tasks that involve distinguishing between different cognitive states or mental activities, such as classifying brain signals associated with motor imagery, attention, or emotional responses. Some of the most notable datasets include:
 % These datasets are crucial for developing and validating deep learning models that can accurately interpret the neural patterns indicative of specific mental tasks, thereby enhancing the capabilities of brain-computer interfaces and cognitive monitoring systems. 
\begin{itemize}
    \item \textbf{BCI Competition IV\cite{Blankertz2007TheNB}}: This dataset comprises multiple sub-datasets, each designed for specific brain-computer interface (BCI) challenges. It includes motor imagery tasks and event-related potentials (ERPs) recorded from healthy subjects.
    \item \textbf{TUH EEG Corpus\cite{Obeid2016TheTU}}: The Temple University Hospital EEG Corpus is one of the largest publicly available EEG datasets. It contains EEG signals collected from 14,987 subjects, with more than 40 different channel configurations and different recording duration, including normal and abnormal samples, making it suitable for both research and clinical applications.
    \item \textbf{DEAP (Database for Emotion Analysis using Physiological Signals\cite{koelstra2011deap})}: This dataset includes EEG and other physiological signals recorded while subjects watched music videos. It is widely used for emotion recognition and affective computing studies.
    \item \textbf{CHB-MIT Scalp EEG Database\cite{shoeb2009application}}: This dataset contains EEG recordings from pediatric subjects with intractable seizures. It is commonly used for seizure detection and prediction research.
    \item \textbf{SEED (SJTU Emotion EEG Dataset)\cite{zheng2017identifying}}: The SEED dataset includes EEG recordings from subjects experiencing emotional stimuli, such as movie clips. It is used to study emotional recognition and related applications.
    % \item \textbf{ZuCo EEG and Eye-Tracking Dataset\cite{Hollenstein2018ZuCoAS}}: This dataset comprises simultaneous EEG and eye-tracking recordings from participants engaged in natural reading and annotation tasks. It is specifically utilized for studying cognitive processes in language understanding and for advancing NLP applications, such as information extraction, entity and relation discovery, and sentiment analysis. 
    \item \textbf{ISRUC-S3 dataset\cite{KHALIGHI2016180}}: This dataset contains 10 healthy subjects. Each recording contains 6 EEG channels, 2 EOG channels, 3 EMG channels, and 1 ECG channel. It is widely used for sleep stage classification studies.
    \item \textbf{MASS-SS3 dataset\cite{2015Montreal}}: This dataset contains 62 healthy subjects. Each recording contains 20 EEG channels, 2 EOG channels, 3 EMG channels, and 1 ECG channel. It is widely used for sleep stage classification studies.
    
\end{itemize}
\textbf{Generative EEG Task Dataset}: These datasets are typically used for tasks that involve the generation of images, sentences, and other signals. For the image generative task, Spampinato et al. \cite{spampinato2017deep}, Kumar et al. \cite{kumar2018envisioned}, and Kaneshiro et al. \cite{kaneshiro2015representational} obtain image semantics from EEG by employing EEG data recorded while subjects looked at images on a screen. The classical dataset constructed for the generative EEG task is shown in Table \ref{tab:eeg_dataset}
\begin{itemize}
    \item Spampinato et al\cite{spampinato2017deep} employed a subset of ImageNet containing 40 classes of easily recognizable objects for visual stimuli, using a 128-channel cap (actiCAP 128Ch), Brainvision DAQs and amplifiers for the EEG data acquisition. Sampling frequency and data resolution were set, respectively, to 1000 Hz and 16 bits. During the recording process, 2,000 images (50 from each class) were shown in bursts for 0.5 seconds each. A burst lasts for 25 seconds, followed by a 10-second pause where a black image was shown for a total running time of 1,400 seconds (23 minutes and 20 seconds).
    \item Kumar et al\cite{kumar2018envisioned} prepared a slide presentation that consisted of 20 text and 10 non-text items in 3 categories of object to the subjects, namely digits, characters and object images, each slide was showed for 10 seconds, then recording the EEG data via a wireless neuro-headset Emotiv EPOC+ at a frequency of 2048Hz and there was a 20 seconds gap between 2 record.
    \item Kaneshiro et al\cite{kaneshiro2015representational} used 72 images from 6 categories of real objects as visual stimuli, acquired the EEG data via 128-channel EGI HCGSN 110 nets in the frequency of 1000 Hz. Each image was displayed for 0.5 seconds, and there was a 0.75 second interval between each image.
    \item \textbf{ZuCo}\cite{hollenstein2018zuco} contains EEG and eyetracking data from 12 healthy adult native English speakers engaged in natural English text reading for 4 - 6 hours. This dataset covers two standard reading tasks and a taskspecific reading task, offering EEG and eye-tracking data for 21,629 words across 1,107 sentences and 154,173 fixations.
\end{itemize}
\subsubsection{\textbf{Private EEG Datasets}}
In addition to publicly available datasets, researchers often collect private EEG datasets tailored to specific research questions or applications. These datasets may focus on particular cognitive tasks, clinical conditions, or subject populations. Specifically, private data also forms the basis of foundation models, and while its importance has been highlighted in \cref{subsec:foundation_model}. Collecting custom datasets allows for greater control over experimental conditions and data quality, but it also requires significant resources and expertise. 
\begin{itemize}
    \item BrainBERT \cite{Wang2023BrainBERTSR} collected stereo electroencephalogram (SEEG) data from 10 subjects(5 male, 5 female; aged 4-19, with a mean age of 11.9 and a standard deviation of 4.6) over 26 sessions, who are pharmacologically intractable epilepsy patients.
    \item BrainNet \cite{Chen2022BrainNet} collected 796 GB of SEEG data from a first-class hospital. The subjects suffering from epilepsy undergo a surgical procedure to implant 4 to 10 invasive electrodes, with 52 to 126 channels, in their brain. In total, the dataset contains 526 hours of 256Hz to 1024Hz recordings.
    \item MBrain \cite{cai2023mbrain} collected 550 GB of SEEG data from a first-class hospital. The subjects suffering from epilepsy undergo a surgical procedure to implant 4 to 10 invasive electrodes, with 52 to 124 channels, in their brain. In total, the dataset contains 470 hours of 1000Hz to 2000Hz recordings.
    \item Brant \cite{Zhang2023BrantFM} collected 1.01 TB of SEEG data from a first-class hospital. The subjects undergo a surgical procedure to implant 4 to 11 invasive electrodes, each with 52 to 153 channels, in their brain. The dataset contains 2528 hours of 1000Hz recordings with more than 1 trillion timestamps. In addition, it also collected 29.39 GB and 43 hours of epilepsy labeled data for fine-tuning of specific downstream tasks.
    \item LaBraM \cite{Jiang2024LargeBM} further collected 342.23 hours of data from more than 140 subjects through the ESI neural scanning system.
\end{itemize}

\subsection{Metrics}

Evaluating the performance of EEG analysis models involves several key metrics, which are crucial for comparing different approaches and understanding their effectiveness. The most commonly used metrics include:
\begin{itemize}
    \item \textbf{Accuracy}: The proportion of correctly classified instances among the total instances. It is a fundamental metric for classification tasks but may be misleading for imbalanced datasets.
    \item \textbf{Precision and Recall}: Precision is the proportion of true positive results among the predicted positives, while recall is the proportion of true positive results among the actual positives. These metrics are particularly useful for tasks with imbalanced classes.
    \item \textbf{F1 Score}: The harmonic mean of precision and recall, providing a single metric that balances both concerns. It is especially useful when the dataset has imbalanced classes.
    \item \textbf{F2 Score}: The harmonic mean of precision and recall, giving twice as much weight to recall. It is particularly useful in applications such as epilepsy detection, where missing positive instances (epileptic events) can be fatal.
    \item \textbf{Area Under the Receiver Operating Characteristic Curve (AUC-ROC)}: This metric evaluates the ability of a model to distinguish between classes, considering both the true positive rate and the false positive rate. It is widely used for binary classification tasks.
    \item \textbf{Mean Squared Error (MSE)}: Used for regression tasks, MSE measures the average squared difference between predicted and actual values. Lower MSE indicates better model performance.
    \item \textbf{Mean Absolute Error (MAE)}: Another metric for regression tasks, MAE measures the average absolute difference between predicted and actual values. It is less sensitive to outliers compared to MSE.
    \item \textbf{Cohen's Kappa}: A statistical measure of inter-rater agreement for categorical items, which takes into account the possibility of agreement occurring by chance. It is useful for evaluating the reliability of classifications.
    % \item \textbf{Confusion Matrix}: A table used to describe the performance of a classification model by displaying the true positives, false positives, true negatives, and false negatives. It provides a comprehensive view of the model's performance.
    \item \textbf{Inception Score (IS)}: A metric used to evaluate the performance of generative models, such as Generative Adversarial Networks (GANs), by assessing the quality and diversity of the generated images. It calculates the classification probabilities of the generated images using a pre-trained Inception network, and measures both how distinct and realistic the generated images are. Higher scores indicate better performance in terms of generating high-quality and diverse images.
    \item \textbf{Frechet Inception Distance (FID)}: A metric for evaluating the quality of generated images by comparing the feature distributions of these images to those of real images. Lower FID scores indicate more realistic and diverse generated images.
    \item \textbf{Kernel Inception Distance (KID)}: A more robust measure of image quality in generative models than FID, KID compares the similarity of feature distributions between generated and real images using a kernel method. It provides a more nuanced assessment by considering both the mean and covariance of the feature distributions, making it sensitive to both the style and content of the images. Lower KID scores suggest better image generation performance.
    \item \textbf{Structural Similarity Index (SSIM)}: A metric for assessing the visual similarity between two images. It evaluates the similarity by comparing the luminance, contrast, and structure of the images. The SSIM index ranges from 0 to 1, with values closer to 1 indicating higher similarity. It is commonly used to measure the effectiveness of image processing techniques like enhancement, compression, and super-resolution.
    \item \textbf{BLEU-N}: A metric used to evaluate the quality of machine-translated text. It measures the correspondence between a machine's translations and human translations by comparing n-gram overlaps. Higher BLEU-N scores indicate better translation accuracy and fluency. BLEU stands for Bilingual Evaluation Understudy.
    \item \textbf{ROUGE-1}: A metric used to evaluate the quality of automatic summarization and machine translation. It focuses on the overlap of unigrams (single words) between a generated summary or translation and a set of reference summaries or translations. Higher ROUGE-1 scores indicate a better match between the generated text and the reference texts.
    \item \textbf{Pearson Correlation Coefficient (PCC)}: a statistical measure that expresses the linear correlation between two variables. It ranges from -1 (perfect negative correlation) to +1 (perfect positive correlation), with 0 indicating no correlation. PCC is commonly used in finance and economics to assess the strength and direction of the relationship between variables.
    \item \textbf{Melcepstral distance}: A measure used in audio processing to evaluate the similarity between two sound signals, often employed in speech recognition and audio analysis. It's calculated based on the Mel-cepstral coefficients derived from the Fourier transform of the audio. Lower melcepstral distances indicate more similar sounds.
    
\end{itemize}

In summary, the availability of diverse and high-quality datasets, combined with robust evaluation metrics, is essential for advancing spatio-temporal EEG data analysis. These resources enable researchers to develop, compare, and refine models, ultimately leading to more accurate and insightful interpretations of brain activity.
\section{Concludes and Future Directions}\label{sec:conclusion}
% \section{Future Research Directions}
\textbf{Conclusion}: In conclusion, this paper has reviewed the current advancements in EEG analysis, focusing on three key areas: representation learning, discriminative-based methods, and generative-based methods. These areas collectively enhance the precision, interpretability, and application scope of EEG signal analysis, addressing significant challenges and paving the way for future research.
\begin{itemize}
    \item \textbf{Learning Useful Representation from EEG Signals}: The first step in understanding EEG signals is representation learning, where we automatically extract important information. Self-supervised learning techniques are effective in this process, helping us create strong representations of EEG signals. These representations improve our ability to interpret the data accurately and handle large amounts of brain signal data efficiently.

    \item \textbf{Identifying Patterns in EEG Signals}: Discriminative methods are crucial for recognizing different patterns or categories within EEG signals. Using advanced techniques like Graph Neural Networks (GNNs) and foundation models, we can gain deeper insights into brain activity by capturing these patterns effectively. Understanding these patterns is essential for deciphering complex neural processes.

    \item \textbf{Generating New Insights from EEG Signals}: Generative methods focus on generating new types of data from EEG signals. Techniques like diffusion models allow us to create images or text based on EEG data, providing innovative ways to visualize and understand brain activity. These generative methods also have applications in generating AI-generated content.
\end{itemize}

\textbf{Future Directions}: Looking ahead, several promising directions for future research in EEG signal analysis and understanding can be identified:

\begin{itemize}
    \item \textbf{Enhanced Integration of Self-Supervised and Semi-Supervised Learning}: Further exploration into the integration of self-supervised and semi-supervised learning techniques could yield even more robust and generalized representations. This will enable better handling of diverse and complex EEG data with minimal labeled data, driving improvements in accuracy and efficiency.

    \item \textbf{Development of Advanced Network Architectures}: Continued innovation in network architectures, such as the refinement and combination of Mamba\cite{gu2023mamba,dao2024transformers}, KAN\cite{liu2024kan}, and MoE models\cite{ni2024mixture}, is essential. These advancements should focus on improving training efficiency and inference speed, particularly for deployment on mobile and edge devices. Research into optimizing these architectures for real-time analysis and low-power consumption is also crucial.

    \item \textbf{Expansion of Multimodal Generative Techniques}: Expanding the capabilities of multimodal generative techniques to include more diverse forms of data, such as tactile or olfactory signals, could open new avenues for EEG applications. Additionally, improving the quality and realism of generated outputs, whether they be images, text, or speech, will enhance their utility in practical scenarios, particularly for assisting individuals with disabilities.

   \item \textbf{Addressing Constrained Conditions in Brain Signals}: Variable missing \cite{yu2024ginar}, class-incremental \cite{qiao2024class}, and source-free domain adaptation \cite{ragab2023source} are constrained conditions in brain signal analysis that present significant challenges but also offer important research opportunities. Addressing these issues can enhance the accuracy and stability of analyses, leading to broad impacts in practical applications.

    \item \textbf{Interdisciplinary Collaboration and Real-World Applications}: Encouraging interdisciplinary collaboration between neuroscientists, computer scientists, and clinicians will be vital for translating these technological advancements into real-world applications. This includes the development of user-friendly interfaces and tools for clinical use, as well as ensuring the ethical and responsible deployment of these technologies.

    % \item \textbf{Personalized and Adaptive EEG Analysis}: Research into personalized and adaptive EEG analysis methods that account for individual variability in brain signals could significantly improve the effectiveness of these techniques. This involves developing models that can dynamically adjust to an individual's unique neural patterns and adapt over time with continuous learning.

    % \item \textbf{Scalability and Large-Scale Data Utilization}: Exploring methods to scale EEG analysis techniques to accommodate large-scale data from diverse populations will be important. This includes addressing challenges related to data heterogeneity, privacy, and the integration of data from various sources to create comprehensive and representative models.
\end{itemize}

By focusing on these future directions, the field of EEG signal analysis can continue to advance, providing deeper insights into brain function and enabling more effective applications in both clinical and non-clinical settings.

% \conflictsofinterest
% The authors declare that they have no conflicts of interest.

\acknowledgement
This work was supported by NSFC grant (No. 62136002), Ministry of Education Research Joint Fund Project(8091B042239), Shanghai Knowledge Service Platform Project (No. ZF1213), and Shanghai Trusted Industry Internet Software Collaborative Innovation Center.
% This work was supported by the NSFC grant (No. 62136002), the Ministry of Education Research Joint Fund Project (8091B042239), the Shanghai Knowledge Service Platform Project (No. ZF1213), and the Shanghai Trusted Industry Internet Software Collaborative Innovation Cen

\bibliographystyle{apalike}
\bibliography{eeg_ref}

\end{document}